# Near-perfect conduction through a ferrocene-based molecular wire

Stephanie A. Getty[1], Chaiwat Engtrakul[2*], Lixin Wang[2], Rui Liu[3], San-Huang Ke[3,4], Harold U. Baranger[4], Weitao Yang[3], Michael. S. Fuhrer[1†], Lawrence R. Sita[2†]

[1]*Department of Physics and Center for Superconductivity Research and* [2]*Department of Chemistry and Biochemistry, University of Maryland, College Park, MD 20742*

[3]*Department of Chemistry and* [4]*Department of Physics, Duke University, Durham, NC 27708*

Here we describe the design, single-molecule transport measurements, and theoretical modeling of a ferrocene-based organometallic molecular wire, whose bias-dependent conductance shows a clear Lorentzian form with magnitude exceeding 70% of the conductance quantum $G_0$. We attribute this unprecedented level of single-molecule conductance to a manifestation of the low-lying molecular resonance and extended orbital network long-predicted for a conjugated organic system. A similar-in-length, all-organic conjugated phenylethynyl oligomer molecular framework shows much lower conductance.

Since its theoretical inception [1], molecular electronics has been an area of significant interdisciplinary interest. The conduction occurring through single molecular states can be manipulated by electric field, light, conformational changes, etc., promising a range of useful devices. The ideal molecular electronic device [2] would allow ballistic conduction between metal reservoirs at the quantum of conductance $G_0 \approx 77$ $\mu$S, corresponding to interface conductivities of order $10^{10}$ S/cm$^2$. To date, delocalized orbital networks in conjugated organic molecules have been proposed as the ideal design for molecular wires. Unfortunately, such all-organic molecules show conductances (in individual molecules) [3-6] and conductivities (in films) [7, 8] orders of magnitude lower than theoretically predicted [9-13].

Our general strategy for the design of components for nanoscale electronic devices has been to incorporate the known favorable physical and electronic properties of ferrocene, ($\eta^5$-C$_5$H$_5$)$_2$Fe, into molecular frameworks [14]. Since recent results have shown that the rate of electron transfer through a conjugated backbone containing a ferrocene moiety can rival that through an all-organic conjugated system [15] and given the proposed relationship between electron-transfer rates and conductance [16], it is of interest to compare the electron transport properties of the two identical-length molecular wires, one containing a ferrocene center [Fc-OPE, Fig. 1(a)] and the other a ferrocene-absent phenylethynyl analog [OPE, Fig. 3(a)].

Synthesis of the molecular adsorbates was achieved in a straightforward manner, although it is important to note that methoxy (OMe) groups were incorporated on the phenyl rings to impart favorable solubility characteristics (rather than for any perceived electronic benefit) [15]. A substrate with a lithographically defined array of (intact) Au wires is immersed in a methylene chloride solution of molecular adsorbates, with or without acetate thiol protecting groups, for approximately 12 hours, whereupon chemisorption of the molecules to the Au occurs.

Upon removal from solution and rinsing, the substrate is wire-bonded and introduced into a sample-in-vapor $^4$He cryostat. Once at base temperature ($T = 1.3$ K), a molecule is localized within a nanometer-sized gap formed via electromigration of a gold wire [17]. The resulting geometry is that of a field effect transistor, with source and drain electrodes bonded to the molecule by the Au-thiol bond. The third gate electrode in our initial work was a degenerately-doped silicon substrate capped with 500 nm of thermally grown $SiO_2$. More recently, we have employed an aluminum layer as a gate electrode and roughly 2 nm of native $Al_2O_3$ as the gate dielectric [18]. More than one device has been measured in each configuration with consistent results.

We first discuss electron transport characteristics at $T = 1.3$ K for five separate gold nanogap junctions formed in the presence of Fc-OPE [Fig. 1(b)-(f)]. As discussed below, we believe each junction contains a single molecule or, in some cases [Fig. 1(c) and (e)], two molecules. Out of over 50 junctions fabricated, only these junctions showed significant conductance; the other junctions showed conductances consistent with tunneling gaps formed in the absence of molecules [19]. For each junction the conductance curves are highly stable and reproducible; this is illustrated in Fig. 1(b) by plotting five sequential sweeps of the bias voltage. Examination of the conductance spectra reveals several common features: (1) a finite conductance at zero source-drain bias $V$, (2) broad resonance peaks occurring at low bias voltage (<100 mV), and (3) a dip in the conductance near $V = 0$ (see insets, Fig. 1). The peak conductances can be a significant fraction of the conductance quantum $G_0$ [60 % in Fig. 1(b), 70% in Fig. 1(d); the conductance greater than $G_0$ in Fig. 1(e) is likely due to two molecules, as discussed below]. The observation of these features only in junctions exposed to Fc-OPE and never in bare nanogaps or nanogaps exposed to OPE (see below) suggests that they are intrinsic to Fc-OPE.

Applied gate voltage, $V_g$, has a variable effect from junction to junction. $V_g$-dependence is shown for the three junctions that exhibit the most pronounced effect [Fig. 1(c), (e), and (f)]. The energy of the lowest-lying molecular level relative to $E_F$ of the electrodes (above or below $E_F$, as determined from the direction of peak motion as a function of $V_g$) also exhibits interjunction variations. This energy distribution is probably due to local electrostatic variations in the presence of an insulating substrate and is not surprising given the closeness of the molecular level to $E_F$, as substantiated by the theoretical calculations discussed below.

Conduction through the ferrocene molecule – or any extended molecular state - should be understood within the theory of resonant conduction through a localized state. The total conductance of a one-dimensional wire is given by the Landauer formula [2]

$$G = \frac{2e^2}{h} \sum_i T_i, \tag{1}$$

where the prefactor is the quantum of conductance, $G_0$, the upper theoretical limit for transmission through a spin-degenerate one-dimensional system, and $T_i$ is the transmission probability for each conducting channel. In the resonant level case, the transmission coefficient is

$$T = \frac{\Gamma_1 \Gamma_2}{\Gamma} \frac{\Gamma}{(E - E_m)^2 + \Gamma^2}, \tag{2}$$

in terms of coupling parameters $\Gamma_1$, $\Gamma_2$, and $\Gamma = \Gamma_1 + \Gamma_2$, which can be modified to include additional lifetime effects, and $E_m$, the energy of the $m$th resonance. We neglect the specific components of the total $\Gamma$ in our analysis and assume that interjunction differences in coupling are dominated by variations in molecule-electrode coupling. We also ignore thermal broadening, which is negligible compared to the experimental peak widths. Ideally, in a symmetric molecule, a single molecular level should result in a pair of Lorentzian peaks in $G(V)$ with half-width $2\Gamma/e$ centered at $\pm 2 E_m/e$, since either forward or reverse bias can bring the resonance into the bias window.

We have performed Lorentzian fits of the observed conductance resonances for two junctions (Fig. 2). We find that the conductance resonances as well as the zero-bias conductance level are well described by the Lorentzian fits (with the exception of the anomalous dip around zero bias). At biases higher than the resonance enhanced conduction is observed, possibly due to additional inelastic channels available for conduction. These fits demonstrate a correlation between high on-resonance conductance and strong molecule-electrode coupling. From Lorentzian fits of all five devices, we have extracted values of $\Gamma$ that vary from 1.2 meV to 12 meV ($\Gamma \gg k_B T$ for all Fc-OPE molecules measured). We note that this Lorentzian resonance is different in nature from the Kondo resonance observed in single-molecule organometallic [18, 20] and fullerene [21] transistors; the Kondo resonance is a collective-electron effect which appears exactly at $E_F$ (zero bias) and is inherently a low-temperature phenomenon. We also observe conductance structure near zero-bias in all five of our ferrocene-based devices which may be due to many-body effects; further studies of this low-bias region are underway.

We now discuss evidence that the observed effects are due to *single* molecules. It is clear that no multiplicative normalization procedure would scale the two curves in Fig. 2 to lie on top of one another. Such an analysis has been proposed to "count" the number of molecules contacted in monolayers [8, 22], however significant differences in bonding and electrostatic environment in few-molecule junctions are likely to cause the observed differences from junction to junction found in our data and in the literature[3-6]: bonding differences are likely to produce changes in the magnitude and width of the resonances[10-12], while electrostatic effects will shift the position of the resonances[23]. In this light, we argue that the observation of a single pair of Lorentzian resonances is the strongest evidence for transport through a single molecule. It is unlikely that a small ensemble of molecules with resonances of different width, height and position could make up the single resonances observed in our data, and more unlikely that each observed resonance would shift uniformly with application of gate voltage, since the gate coupling should vary significantly from molecule to molecule.

In some cases two pairs of resonances are resolved in the devices (Fig. 1(c) and (e)). An applied gate voltage has a separable effect on each pair of resonances; the effect is particularly striking in Fig. 1(e) where one molecular resonance can actually be seen to move through the other. The independent behavior of the two pairs of resonances indicates that they are not from two levels of the same molecule but rather each arises from a different molecule. The separable effect suggests that the two molecules are sufficiently spaced from each other within the nanogap to be differently coupled to the gate electrode. The presence of two molecules is likely responsible for conductance that exceeds $G_0$ in Fig. 1(e).

To investigate the influence of the ferrocene center on the molecular transport, we have also measured the electron transport through the all-organic molecule OPE [Fig. 3(a)]. Following identical sample preparation as for Fc-OPE, very different behavior was observed for this conjugated species. Current-voltage measurements are shown [Fig. 3(b)-(e)] for four molecular junctions out of approximately 50 nanogaps measured. The on-resonance conductance, obtained by numerical differentiation, is in the nS range, at least two orders of magnitude lower than in the ferrocene-based species. A zero-conductance gap is observed over several hundred mV, similar to what has been seen by other groups in single conjugated organic molecule junctions [4, 5]. The Fc-OPE data are replotted in Fig. 3(b) to emphasize the differences between these two molecules.

Theoretical calculations support the presence of a low-lying, high-transmission molecular level in the Fc-OPE molecule. We use density functional theory (DFT) for the electronic structure combined with Green function techniques to calculate the transmission at zero bias [2, 9-13, 19, 24]. For the optimized planar Fc-OPE [Fig. 4(a)], *T(E)* shows a clear resonance 30 meV above the Fermi energy [Fig. 4(c)], attaining nearly perfect transmission on resonance. The local density of states at the energy of the peak [Fig. 4(b)] shows a conjugated level traversing the entire molecule from lead to lead: this is the molecular level causing the high conductance seen experimentally. Turning to the OPE control molecule, we find that for the configuration in which all five rings are co-planar, the conductance is high, in sharp disagreement with the experiments. This is an example of the well-known discrepancy between DFT [9-13] and experiment [3-8] for all-organic conjugated molecules. While our work does not solve this long-standing problem, we emphasize that our measurements of the Fc-OPE molecule constitute the first instance where the simple theoretical scenario of resonant conduction through a single molecular level has been realized.

The precise role of the ferrocene in this regard is, however, not clear. One possibility is that it enhances the co-planarity of the rings; certainly, the conductance of an OPE molecule in a non-co-planar configuration is very low because of the broken conjugation. Another possibility is that the scissor mode made possible by the ferrocene moiety (i.e. rotation of the two five-member rings with respect to each other) allows a high conductance state to be realized – our calculations show that a bent Fc-OPE has a conductance lower than the linear one but within the experimental variation. Experimentally, the low conductance arrangement for OPE may be favored by steric repulsion or hydrogen bond formation caused by the methoxy side groups.

We have demonstrated the chemical modification of a simple conjugated molecular system to create an excellent molecular wire out of an otherwise poor (all-organic) conductor through incorporation of a central ferrocene unit. We believe that the ferrocene moiety is an attractive basis for a device "toolbox", in which a spectrum of device properties could be achieved by varying the ferrocene number, symmetry, and functionality of the linker segments. The molecule presently under study is a simple precursor to more complex chemically engineered species.

The authors are grateful for support by the DCI Postdoctoral Fellowship Program, the National Science Foundation (to H. U. B., W. Y., M. S. F., and L. R. S.), the Department of Energy (to M. S. F. and L. R. S.) and the NSF-MRSEC at the University of Maryland.

*Present address: National Renewable Energy Laboratory, Golden, CO 80401

†Corresponding authors. Email: mfuhrer@umd.edu, lsita@umd.edu

FIG. 1. (a) Highly conducting Fc-OPE, approximately 3.5 nm in length, featuring methoxy side groups [14]. (b)-(f) Current $I$ (red) in nA and conductance $G$ (blue) in units of $G_0$ versus source-drain voltage $V$ in Fc-OPE at low temperature ($T < 1.5$ K) for five separately formed molecular junctions. Low-energy structure on the 5-10 mV scale is shown in the insets. In (b), five separate sweeps from $V = -100$ mV to 100 mV are shown; the curves are nearly indistinguishable. (b)-(d) were studied using a doped Si backgate and 500-nm $SiO_2$ dielectric, using a solution of the with free terminal thiol (-SH) groups that are known to readily form Au-S bonds. (e) and (f) were studied using an Al/native-$Al_2O_3$ near-by gate configuration, using a solution of thioacetate-protected adsorbates. As similar results have been obtained for both types of solutions, formation of Au-S bonds can occur following apparent Au-catalyzed *in situ* acetate group removal. Effectiveness of the gate voltage $V_g$ is most pronounced in (c), (e), and (f); we attribute the irreproducibility of peak position to variations in the local electrostatic environment, giving rise to an effective offset in $V_g$. The $V_g$-dependences are shown in color, with dark blue being the most negative and magenta being the most positive: (c) -5 V $\leq V_g \leq$ 5 V; (e) -1 V $\leq V_g \leq$ 1 V; (f) 1.1 V $\leq V_g \leq$ 1.7 V, evenly spaced in $V_g$. (c) and (e) each exhibit two pairs of conductance peaks that respond independently to $V_g$, indicating that these junctions each contain two molecules. Diamond and star shapes, colored to denote varying $V_g$, illustrate the motion of each peak, best resolved at negative $V$.

FIG. 2. $G$ in units of $G_0$ versus $V$ (mV) for two separately formed Fc-OPE molecular junctions. The red curve represents a highly conducting single molecule from Fig. 1(b); the blue curve a moderately conducting junction consisting of two molecules from Fig. 1(c). We interpret each pair of resonances in the blue curve to represent a single molecule. The thin black curves are Lorentzian fits of the data, using two peaks and four peaks for fitting the red curve and blue curve, respectively.

FIG. 3. (a) Poorly conducting non-ferrocene containing phenylethynyl dithiol (OPE), featuring methoxy side groups for solubility purposes. (b)-(e) $I$ (nA) versus $V$ (mV) in the ferrocene-less control species at $T < 1.5$ K. For comparison, the *I-V* curves for the five Fc-OPE junctions, Fig. 1 (b)-(f), are reproduced in (b) (black lines). The four OPE curves demonstrate the relative suppression of current levels and the development of a zero-conductance gap in the absence of the ferrocene moiety. As in the Fc-OPE, the effectiveness of the gate is seen to vary between junctions. All devices were measured in the Si/SiO$_2$ global gate geometry; variations in $V_g$ are denoted in the figure legends.

FIG. 4. (a) Relaxed configuration of Fc-OPE between two gold (001) leads. All the phenyl rings are coplanar. (b) Surface of constant local density of states (LDOS) at the energy of peak transmission. (c) Transmission between the gold leads (dashed) and the density of states projected onto the molecule (solid) at zero bias. Note the resonance 30 meV above the Fermi energy in both curves. The molecular level causing the large transmission is imaged through the LDOS in panel (b).

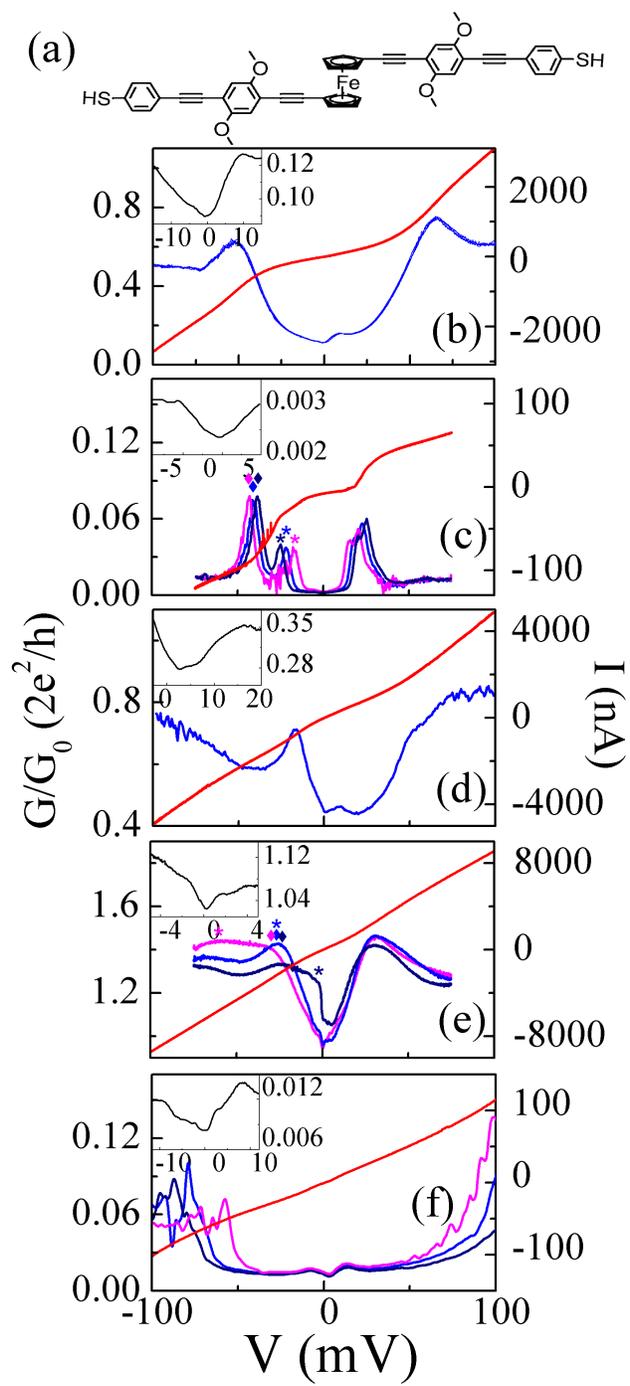

FIG. 1

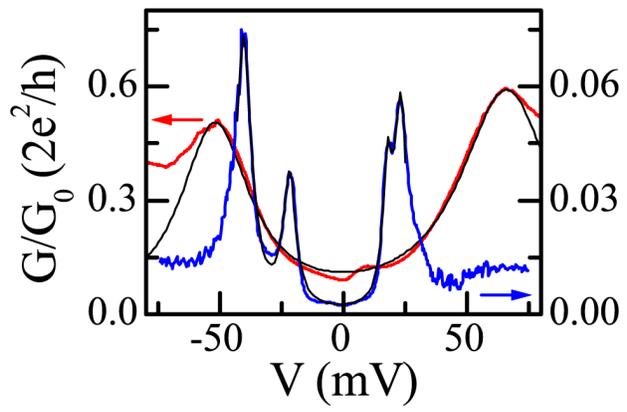

FIG. 2

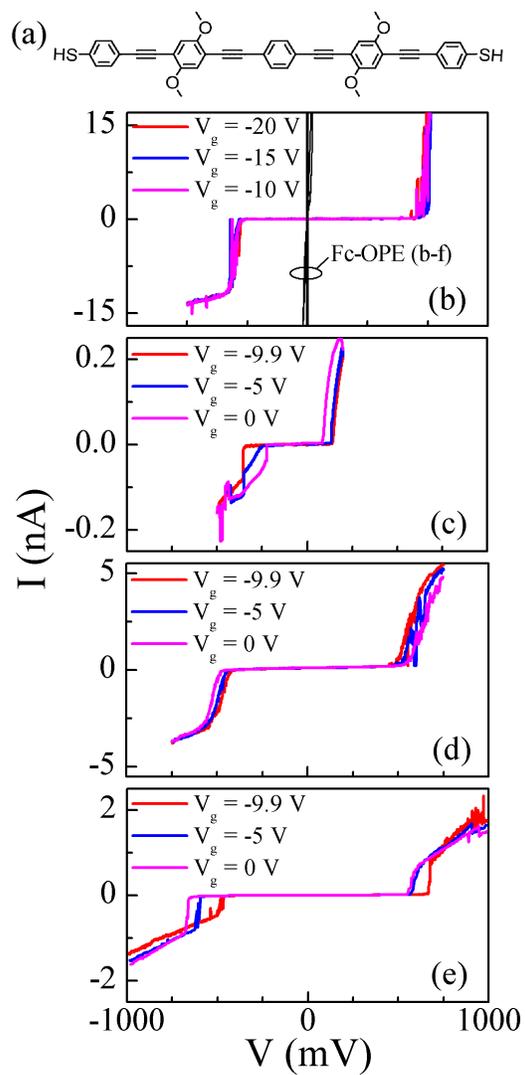

FIG. 3

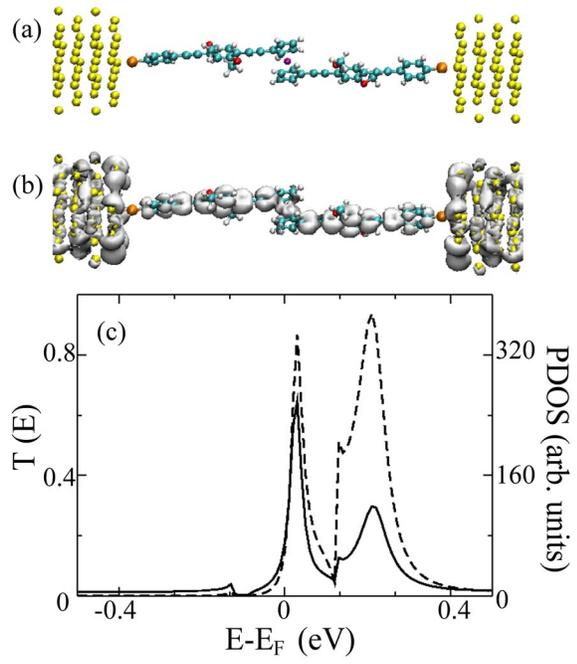

FIG. 4

# Supplementary Information

Here we address three issues related to "Near-perfect conduction through a ferrocene-based molecular wire," an experimental and theoretical study of the enhancement of molecular conductance through a conjugated species upon incorporation of ferrocene. The first section details to the reader the characterization of empty nanogaps prior to, and as a comparison for, the incorporation of molecular adsorbates. The second section discusses the relevance of Coulomb blockade to the understanding of transport through a ferrocene-containing conjugated species and its ferrocene-less analogue. The third section discusses some aspects of the theoretical calculations in greater detail.

## 1. CONTROL MEASUREMENTS

We have characterized over 75 nanogaps formed in the absence of any molecules. Typical transport characteristics exhibit nS- to pS-scale conductance levels, and a positive curvature in the conductance as a function of voltage that is consistent with tunneling between two metal reservoirs. We have also occasionally observed nanogaps that exhibit Coulomb blockade with characteristic charging energies of a few meV, in which numerous equally-spaced charge states can be accessed easily with gate voltage. We ascribe such features to formation of an accidental single-electron transistor consisting of a gold island inside of or near to the nanogap.

## 2. THE ROLE OF COULOMB BLOCKADE

The differences in behavior of these two molecular species, i.e. ferrocene and non-ferrocene containing, cannot be explained within the Coulomb blockade model. For example, if we associate the separation of the resonances in the ferrocene-based molecule with a Coulomb charging energy $E_c \sim 50$ meV, we expect a temperature of 1.3 K to be sufficiently low for complete gap formation. Within the gap, then, we would expect the conductance to be suppressed exponentially by a factor $\exp(-E_c/k_B T) \sim \exp(-500)$, where $k_B T$ is the thermal energy, approximately 110 $\mu$eV in our experiment. Instead, we observe residual intra-gap conductance that reproducibly and generically features an anomalous dip at zero bias, discussed below. Furthermore, in the quantum Coulomb blockade regime, the conductance suppression occurs over an energy gap given by

$$E_G = E_c + \Delta E,  \qquad (1)$$

where $E_c$ is the Coulomb charging energy, and here, $\Delta E$ is the energy difference between the lowest-lying molecular level and $E_F$ of the electrodes [1]. We recall that, for sufficient screening within the quantum system, $E_c$ is a geometrical quantity, and as such, one would expect a similar gap to occur in both the ferrocene and non-ferrocene containing molecular frameworks. Experimentally, however, we find the average gap value (at $V_g = 0$ V) in the ferrocene-containing species to be 49 meV, and that of the non-ferrocene species to be 440 meV. The disagreement between these two values suggests that $E_G$ is dominated by $\Delta E$ and that Coulomb blockade is a negligible effect.

## 3. THEORETICAL METHODS

The current method of choice for calculating electron transport through molecules attached to leads is density functional theory (DFT) for the electronic structure combined with non-equilibrium Green function (NEGF) theory for the quantum transport [2-9]. Our own implementation of this method is described in detail in Ref. 9; here we provide a brief summary.

For the electronic structure calculation, we use Siesta, an efficient full DFT package [10]. The atomic structure of the molecule and the molecule-lead separation are fully optimized, with residual forces less than 0.02 eV/Å, while the atoms in the leads are fixed at their bulk positions. Each lead presents a (001) surface to the molecule and is 4√2 by 4√2 in size; the molecule is adsorbed at the hollow site. Relaxation of the gold lead was performed for smaller molecules (1 phenyl ring) and shown to be small for this type of lead [11]. The PBE version of the generalized gradient approximation (GGA) is used for the electron exchange and correlation [12], and optimized Troullier-Martins pseudopotentials [13] are used for the atomic cores. We used a high-level double-zeta plus polarization basis set (DZP) in relaxing the molecules and a single-zeta plus polarization basis set (SZP) for the transmission calculation. We have compared the use of SZP to DZP for transmission through a single dithiolated phenyl ring between Au leads. These studies suggest that an SZP treatment, while not quantitatively precise, is sufficiently accurate to capture the main features and magnitudes reported here.

For the transport calculation [9] we divide an infinite lead-molecule-lead system into three parts: left lead, right lead, and device region which contains the molecule and large parts of the left and right leads so that the molecule-lead interactions can be fully accommodated. For a steady state situation in which the device region is under a bias $V_b$ (zero or finite), its density matrix and Hamiltonian can be determined self-consistently by the DFT+NEGF method. The Kohn-Sham wave-functions are then used to construct a single-particle Green function from which the transmission coefficient at any energy, $T(E)$, is calculated. The conductance, $G$, then follows from a Landauer-type relation.